\documentclass{article}
\usepackage{spconf,amsmath,graphicx, enumitem,booktabs,tabto,xcolor,multirow,bbding,breakurl}
\usepackage{amsfonts}
\usepackage{hyperref}

\title{ReFlow-TTS: A Rectified Flow Model for High-fidelity Text-to-Speech}
\name{Wenhao Guan$^1$, Qi Su$^2$, Haodong Zhou$^2$, Shiyu Miao$^2$, Xingjia Xie$^2$, Lin Li$^{*2}$, Qingyang Hong$^{*1}$
\thanks{\quad $^{*}$ Corresponding author.}
\thanks{This work was supported in part by the National Natural Science Foundation of China under Grants 62001405, 62276220, and 61876160.
}}
\address{
  $^1$School of Informatics, Xiamen University, China\\
  $^2$School of Electronic Science and Engineering, Xiamen University, China\\
  }

\begin{document}
\ninept
%
\maketitle

\begin{abstract}
The diffusion models including Denoising Diffusion Probabilistic Models (DDPM) and score-based generative models have demonstrated excellent performance in speech synthesis tasks. However, its effectiveness comes at the cost of numerous sampling steps, resulting in prolonged sampling time required to synthesize high-quality speech. This drawback hinders its practical applicability in real-world scenarios. In this paper, we introduce ReFlow-TTS, a novel rectified flow based method for speech synthesis with high-fidelity. Specifically, our ReFlow-TTS is simply an Ordinary Differential Equation (ODE) model that transports Gaussian distribution to the ground-truth Mel-spectrogram distribution by straight line paths as much as possible.
Furthermore, our proposed approach enables high-quality speech synthesis with a single sampling step and eliminates the need for training a teacher model. Our experiments on LJSpeech Dataset show that our ReFlow-TTS method achieves the best performance compared with other diffusion based models. And the ReFlow-TTS with one step sampling achieves competitive performance compared with existing one-step TTS models.
\end{abstract}
\section{Introduction}
\label{sec:intro}
Speech synthesis, also known as Text-to-Speech (TTS), is a field that focuses on generating natural and intelligible speech from textual input. It plays a crucial role in various applications, such as voice assistants, audiobooks, and accessibility tools.
With the advancements in deep learning, significant strides have been made in the field of speech synthesis. Presently, the majority of state-of-the-art neural speech synthesis systems employ a two-stage pipeline including an acoustic model and a vocoder.
In a two-stage speech synthesis system, the acoustic model serves as the first stage, transforming textual information into Mel-spectrograms. Subsequently, the vocoder converts the generated Mel-spectrograms into speech waveforms. It is worth noting that the quality of the synthesized speech mainly relies on the acoustic features produced by the acoustic models.

The typical speech synthesis models, such as Tacotron \cite{wang17n_interspeech}, Transformer-TTS~\cite{li2019neural},  FastSpeech \cite{ren2019fastspeech}, DurIAN \cite{yu20c_interspeech}, have achieved tremendous success. But there is still room for improvement of the acoustic model. The diffusion models including the denoising diffusion probabilistic models (DDPM)~\cite{ho2020denoising} and score-based generative models~\cite{song2020score} have attracted much attention due to its potential to generate high-quality samples.

However, a notable drawback of diffusion models is its reliance on many iterations to generate satisfactory samples.  
And several methods have been proposed for speech synthesis utilizing diffusion models~\cite{jeong21_interspeech,liu2022diffsinger,popov2021grad,guan23_interspeech}. Nonetheless, a prevalent challenge encountered by these methods is the issue of slow generation speed.

Diff-TTS \cite{jeong21_interspeech} leverages a DDPM framework to convert a noise signal into a Mel-spectrogram through multiple diffusion time steps.
DiffSpeech \cite{liu2022diffsinger} introduces a shallow diffusion mechanism to enhance voice quality and accelerate inference speed.
Grad-TTS \cite{popov2021grad} formulates a stochastic differential equation (SDE) to gradually transform noise into a mel-spectrogram, employing a numerical ODE solver to solve the reverse SDE. Although it produces high-quality audio, the inference speed is slowed down due to the large number of iterations in the reverse process.
ProDiff \cite{huang2022prodiff} further develops the approach by utilizing progressive distillation to reduce the number of sampling steps.
DiffGAN-TTS \cite{liu2022diffgan} adopts an adversarially-trained model to approximate the denoising function, enabling efficient speech synthesis.
ResGrad \cite{chen2022resgrad} employs the diffusion model to estimate the prediction residual from a pre-trained FastSpeech2 \cite{ren2020fastspeech} model and ground truth.
CoMoSpeech \cite{ye2023comospeech} achieves one-step high-quality speech synthesis by utilizing a consistency model~\cite{song2023consistency}, which is distilled from a pre-trained teacher model.
A recent work Voicebox~\cite{le2023voicebox} is proposed to generate masked speech given its surrounding audio and text transcript on a text-guided speech infilling task, which is a non-autoregressive continuous normalizing flow (CNF) model~\cite{chen2018neural} trained with flow-matching~\cite{lipman2022flow} method.

\begin{figure*}[!t]
  \centering
  \includegraphics[width=0.91\linewidth]{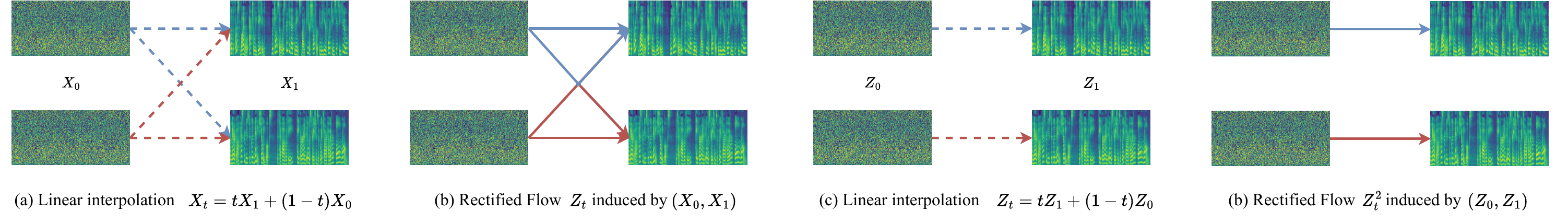}
  \caption{The graphical model for the Rectified Flow Model. (a) Linear interpolation of data samples $(X_{0},X_{1})$. (b) The Rectified Flow $Z_{t}$ induced by $(X_{0},X_{1})$. (c) The linear interpolation of data samples $(Z_{0},Z_{1})$ of rectified flow $Z_{t}$. (d) The rectified flow induced from $(Z_{0},Z_{1})$ and it follows straight paths.}
  \label{fig:reflow}
\end{figure*}

In this paper, we introduce ReFlow-TTS, a speech synthesis model based on the Rectified Flow model~\cite{liu2022flow}. Our proposed approach achieves exceptional speech synthesis results, and surpasses the performance of most diffusion based TTS models with just one sampling step during the inference stage. Notably, our model eliminates the need for pre-training a teacher model, further streamlining the synthesis process compared with CoMoSpeech. The contributions of our work are as follows:
\begin{itemize}
\item We propose ReFlow-TTS, which is the first acoustic model for TTS based on the Rectified Flow model. Specifically, the ReFlow-TTS model is an ODE model that transports Gaussian distribution to the ground-truth Mel-spectrogram distribution by straight line paths as much as possible. And it is trained with a simple unconstrained least squares optimization procedure.
The ReFlow-TTS can generate high-fidelity speech samples using a numerical ODE solver.
\item We show that ReFlow-TTS can achieve competitive performance compared to most existing diffusion based speech synthesis models with only one sampling step in the inference stage, and it does not rely on pre-training a teacher model.

\end{itemize}

\vspace{-5pt}
\section{Rectified Flow Model}
\vspace{-5pt}
In this section, we introduce the background of rectified flow model~\cite{liu2022flow}.
The rectified flow model is an ODE model that transports distribution $\pi_{0}$ to $\pi_{1}$ by straight line paths as much as possible, where $\pi_{0}$ is the standard Gaussian distribution and $\pi_{1}$ is the ground truth distribution. 

\textbf{Overview}
Given empirical observations of $X_{0}\sim\pi_{0}$ and $X_{1}\sim\pi_{1}$, the rectified flow induced from $(X_{0},X_{1})$ corresponds to an Ordinary Differential Equation (ODE) with respect to time $t \in [0,1]$. This ODE can be represented as:
\begin{equation}
dZ_{t}=v(Z_{t},t)dt,
\end{equation}
where $Z_{0}$ from the distribution $\pi_{0}$ is transformed to $Z_{1}$ following the distribution $\pi_{1}$. $v$ is the drift force of the ODE, which is designed to drive the flow in a manner that aligns with the direction $(X_{1}-X_{0})$ of the linear path connecting $X_{0}$ and $X_{1}$. This mapping is achieved by solving a simple least squares regression problem:
\begin{equation}
\label{maineq}
\mathop{\min}\limits_{v}\int_{0}^{1} \mathbb{E} [||(X_{1}-X_{0})-v(X_{t},t)||^{2}]dt,
\end{equation}
where $X_{t}=tX_{1}+(1-t)X_{0}$, $X_{t}$ is the linear interpolation of $X_{0}$ and $X_{1}$.

Naively, the evolution of $X_{t}$ follows the ODE $dX_{t}=(X_{1}-X_{0})dt$, which is non-causal because it introduces a dependency on the final point $X_{1}$ for updating $X_{t}$. However, by adjusting the drift force $v$ based on the difference $(X_{1}-X_{0})$, the rectified flow causalizes the paths of linear interpolation $X_{t}$, which allows for simulating the rectified flow without requiring knowledge of future states.

A key aspect of understanding the method lies in the non-crossing property of flows. When following a well-defined ODE, expressed as $dZ_{t}=v(Z_{t},t)dt$, where the solution is unique and solvable, the different paths cannot cross each other at any point in time $t \in [0,1]$. In other words, there is no location $z\in \mathbb{R}^{d}$ and time $t\in [0,1]$ where two paths intersect at $z$ along different directions. If such crossings were to occur, the solution of the ODE would not be unique.

However, in the case of the interpolation process $X_{t}$, the paths may intersect with each other (as depicted in Figure \ref{fig:reflow} (a)), making it non-causal. To address this, the rectified flow adjusts the individual trajectories passing through the intersection points to avoid crossing, while simultaneously tracing out the same density map as the linear interpolation paths. This alignment is achieved through the optimization of equation \eqref{maineq}.
On the other hand, the rectified flow can be seen as the traffic of particles passing through these roads in a  non-crossing manner. This allows the particles to disregard the global path information regarding the pairing of $X_{0}$ and $X_{1}$, and instead establish a more deterministic pairing of $(Z_{0},Z_{1})$. This process is illustrated in Figure \ref{fig:reflow} (b).

\textbf{Training}
For training the Rectified Flow model, we solve the equation \eqref{maineq} to learn the parameter $\theta$. 

We first prepare the samples from $(X_{0},X_{1})$ of $\pi_{0}$ and $\pi_{1}$, and a drift force model $v_{\theta}$. The training objective is as follows:
\begin{equation}
\label{reflow_train}
\hat{\theta}=\mathop{\arg\min}\limits_{\theta} \mathbb{E} [||(X_{1}-X_{0})-v(X_{t},t)||^{2}],
\end{equation}
where $t \sim Uniform([0,1])$ and $\hat{\theta}$ is the learned optimal parameter. 
After training, we get $v$ following $dZ_{t}=v_{\hat{\theta}}(Z_{t},t)dt$, and we solve the ODE starting from $X_{0} \sim \pi_{0}$ to transfer $\pi_{0}$ to $\pi_{1}$ for sampling.

We define the procedure as $Z=Reflow((X_{0},X_{1}))$. Applying this procedure recursively yields a new second rectified flow $Z^{2}=Reflow((Z_{0},Z_{1}))$, where $Z_{0}$ is the samples from Gaussian distribution and $Z_{1}$ is the generated samples from the procedure $Z=Reflow((X_{0},X_{1}))$. 
The recursive rectified flow procedure serves a dual purpose of reducing transport costs and straightening the paths of rectified flows, leading to a more linear flow trajectory. This computational advantage is particularly valuable as it minimizes time-discretization errors when numerically simulating flows with almost straight paths.

\begin{figure}[!t]
  \centering
  \includegraphics[width=0.83\linewidth]{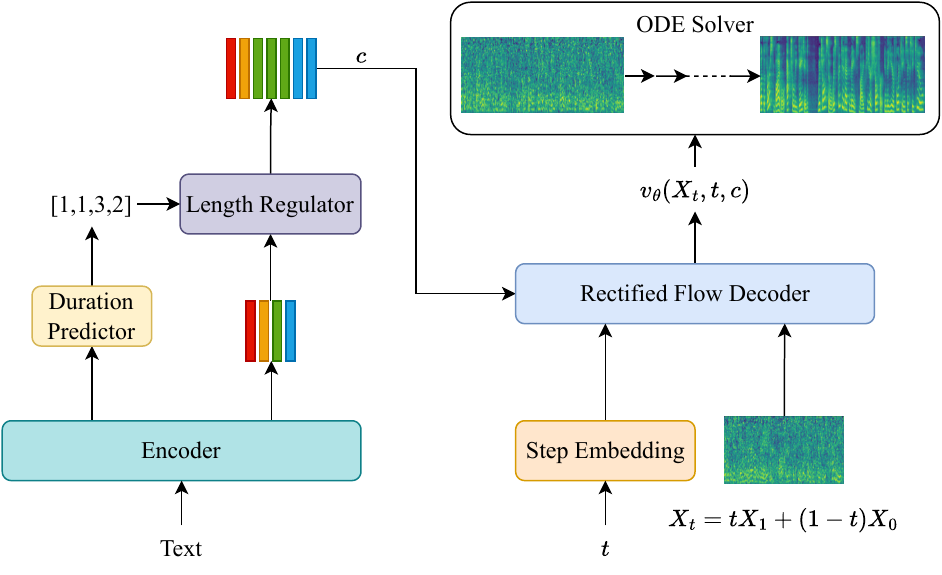}
  \caption{An illustration of ReFlow-TTS.}
  \label{fig:reflow-tts}
\end{figure}
\section{ReFlow-TTS}
In this section, we first present the rectified flow model of ReFlow-TTS for Mel-spectrogram generation. And then we present the model architecture of ReFlow-TTS.

\begin{figure*}[!t]
  \centering
  \includegraphics[width=0.93\linewidth]{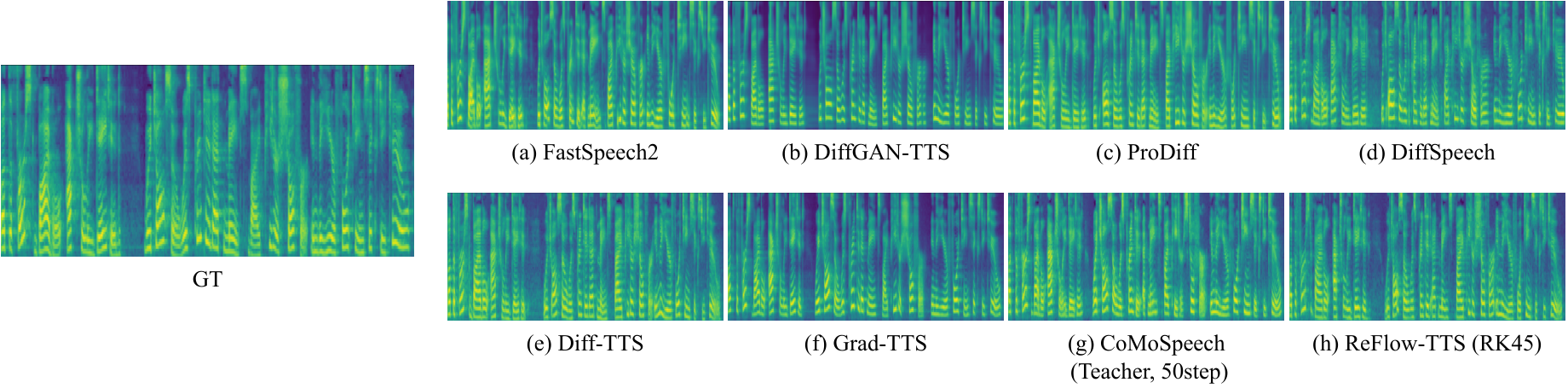}
  \caption{The visualization results of Mel-spectrograms for compared models.}
  \label{fig:comp_models}
\end{figure*}

\subsection{Rectified Flow Model for TTS}
ReFlow-TTS converts the noise distribution to a Mel spectrogram distribution conditioned on time $t$ and text condition feature $c$. We define the $\pi_{0}$ as the standard Gaussian distribution and $\pi_{1}$ as the ground truth Mel-spectrogram data distribution, and $X_{0} \sim \pi_{0}$, $X_{1} \sim \pi_{1}$. 
The training objective of ReFlow-TTS is as follows:
\begin{equation}
\label{reflow_train}
L_{\theta}=\mathbb{E} [||(X_{1}-X_{0})-v_{\theta}(X_{t},t,c)||^{2}],
\end{equation}
where $t \in Uniform([0,1])$ and $X_{t}=tX_{1}+(1-t)X_{0}$. 
ReFlow-TTS does not need any auxiliary losses except L2 loss function between the output of model $v_{\theta}$ and $(X_{1}-X_{0})$.

During inference phase, we directly solve the ODE starting from $Z_{0} \sim \pi_{0}$ conditioned on the text feature $c$ and based on model $v_{\theta}$. In this work, we can use the RK45 ODE solver for high-fidelity generation. For one-step generation, we can directly use Euler ODE solver for competitive performance.

Furthermore, the recursive rectified flow procedure can also be applied in TTS to construct a second ReFlow-TTS, which is named as 2-ReFlow-TTS. The 2-ReFlow-TTS is simply re-train the rectified flow model using the generated samples by ReFlow-TTS.

\begin{figure}[!t]
  \centering
  \includegraphics[width=0.99\linewidth]{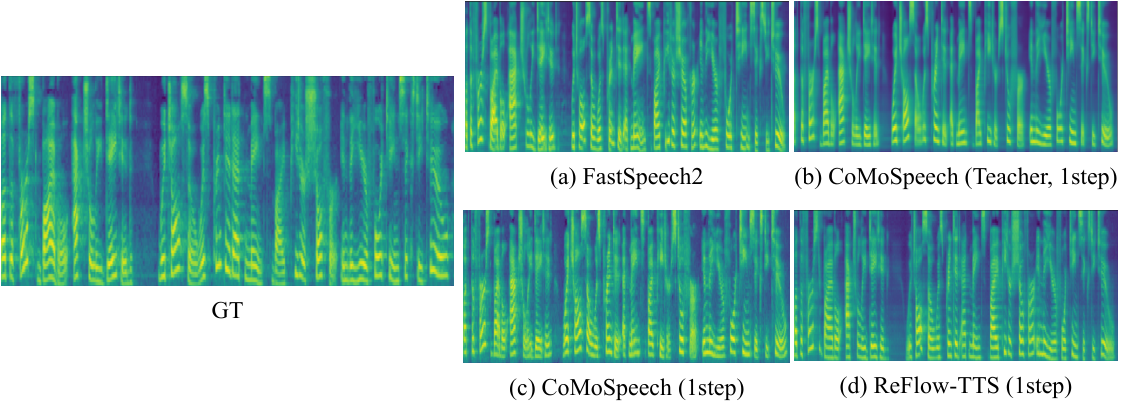}
  \caption{The visualization results of Mel-spectrograms for one step sampling TTS models.}
  \label{fig:exp_1step}
\end{figure}

\begin{table}[!t]
	  \centering  
	  \caption{Evaluation results on LJSpeech for TTS.}
	  \label{t1} 
	  \begin{center}
	   \resizebox{8.5cm}{!}{
		\begin{tabular}{lcccc}\hline 
 		Method & NFE $\downarrow$ & RTF $\downarrow$ & FD $\downarrow$ & MOS $\uparrow$  \\ \hline
 		GT & - & - & - & 4.69$\pm$0.10 \\
 		GT (Mel+Voc) & - & - & 0.1121 & 4.57$\pm$0.09 \\  \hline
 		FastSpeech2~\cite{ren2020fastspeech} & \textbf{1} & \textbf{0.00293} & 4.2607 & 3.73$\pm$0.11 \\
 		DiffGAN-TTS~\cite{liu2022diffgan} & 4 & 0.00856 & 5.1098 & 3.81$\pm$0.09 \\
 		ProDiff~\cite{huang2022prodiff} & 4 & 0.01276 & 1.8835 & 3.69$\pm$0.10 \\
 		DiffSpeech~\cite{liu2022diffsinger} & 71 & 0.3728 & 0.9864 & 4.12$\pm$0.07 \\
 		Grad-TTS~\cite{popov2021grad} & 50 & 0.3185 & 0.4186 & 4.26$\pm$0.09 \\
 		CoMoSpeech (Teacher, 50step)~\cite{ye2023comospeech} & 50 & 0.2256 & 0.6911 & 4.33$\pm$0.10 \\
 		Diff-TTS~\cite{jeong21_interspeech} & 1000 & 1.2639 & \underline{0.2182} & \underline{4.51$\pm$0.11} \\ \hline
 		ReFlow-TTS (1step)  & 1 & \underline{0.00577} & 1.9872 & 4.16$\pm$0.09 \\
 		ReFlow-TTS (50 step) & 50 & 0.0964 & 0.2759 & 4.43$\pm$0.07 \\
 		ReFlow-TTS (RK45 solver) & 152 & 0.3703 & \textbf{0.1393} & \textbf{4.52$\pm$0.10} \\
 		\hline
		\end{tabular}}
	\end{center}
\end{table}

\subsection{Model Architecture}
For the model structure of ReFlow-TTS, it consists of  text encoder, step encoder, duration predictor, length regulator and rectified flow decoder. The architecture of ReFlow-TTS is illustrated as Figure \ref{fig:reflow-tts}.

The encoder, duration predictor, and length regulator employed in this work follow a similar setup as described in FastSpeech2~\cite{ren2020fastspeech}. Specifically, the encoder module is responsible for encoding the input text into linguistic hidden features. The length regulator module is utilized to expand the linguistic hidden features to match the length of the corresponding Mel-spectrograms, which is determined based on the duration information extracted by the duration predictor.

The step encoder converts the step $t$ to a step embedding using the sinusoidal position embedding~\cite{vaswani2017attention} with 256 channels.  For the rectified flow decoder, we adopt a similar architecture as in DiffWave~\cite{kong2020diffwave}.  The decoder network comprises a stack of 20 residual blocks incorporating Conv1D, tanh, sigmoid and 1x1 convolutions with 256 channels.

\section{Experiments}
\subsection{Experimental Setup}
\subsubsection{Dataset}
We conducted an evaluation of the proposed ReFlow-TTS using the LJSpeech dataset~\cite{ljspeech17}, which comprises recordings sampled at
22.05kHz from a single female speaker. The dataset consists of 13,100 speech samples, approximately 24 hours in duration. The dataset was randomly split into the training set (12,500 samples), validation set (100 samples), and test set (500 samples). We extract the 80-bin mel-spectrogram with the frame size of 1024 and hop size of 256.
The ReFlow-TTS was trained for 300K iterations using Adam optimizer~\cite{kingma2014adam} on a single NVIDIA 2080Ti GPU. We employed a pre-trained HiFi-GAN~\cite{kong2020hifi} as a neural vocoder, responsible for converting the Mel-spectrogram into the raw waveform.

\vspace{-5pt}
\subsubsection{Evaluation Metrics}
\vspace{-5pt}
We conducted a comprehensive evaluation, encompassing both objective and subjective measures, to assess the sample quality (MOS and FD) as well as the model's inference speed (NFE and RTF).  We employ MOS (mean opinion score) with 95\% confidence interval to assess the perceived quality of synthesized audio and a test set is presented to 10 listeners who rate the audio quality on a 1-5 scale. The Frechet Distance (FD) is similar to the frechet inception distance in image generation. We use frechet distance~\cite{liu2023audioldm} in audio to measure the similarity between generated samples and target samples. The Number of Function Evaluations (NFE) measures the computational cost by tracking the total number of evaluations made to the decoder function during the generation process. The Real-Time Factor (RTF) determines the real-time synthesis capability of a system. It is calculated as the ratio between the total synthesis time for a given audio and the duration of that audio. We obtain the RTF through a single NVIDIA 2080Ti GPU.

\begin{table*}[!t]
	  \centering  
	  \caption{Evaluation results for TTS models using one sampling step.}
	  \label{t2} 
	  \begin{center}
	   \resizebox{13cm}{!}{
		\begin{tabular}{lccccc}\hline 
 		Method & Teacher Pre-training & NFE $\downarrow$ & RTF $\downarrow$ & FD $\downarrow$ & MOS $\uparrow$  \\ \hline
 		FastSpeech2~\cite{ren2020fastspeech} & No & 1 & \textbf{0.00293} & 4.2607 & 3.73$\pm$0.11 \\
 		CoMoSpeech (Teacher, 1step)~\cite{ye2023comospeech} & No & 1 & 0.00609 & 4.3696 & 3.36$\pm$0.10 \\
 		CoMoSpeech ~\cite{ye2023comospeech} & Yes & 1 & 0.00606 & \textbf{0.8179} & \textbf{4.21$\pm$0.09} \\
 		ReFlow-TTS (1step) & \textbf{No} & 1 & \underline{0.00577} & \underline{1.9872} & \underline{4.16$\pm$0.09} \\
 		\hline
		\end{tabular}}
	\end{center}
\end{table*}

\begin{figure*}[!t]
  \centering
  \includegraphics[width=0.81\linewidth]{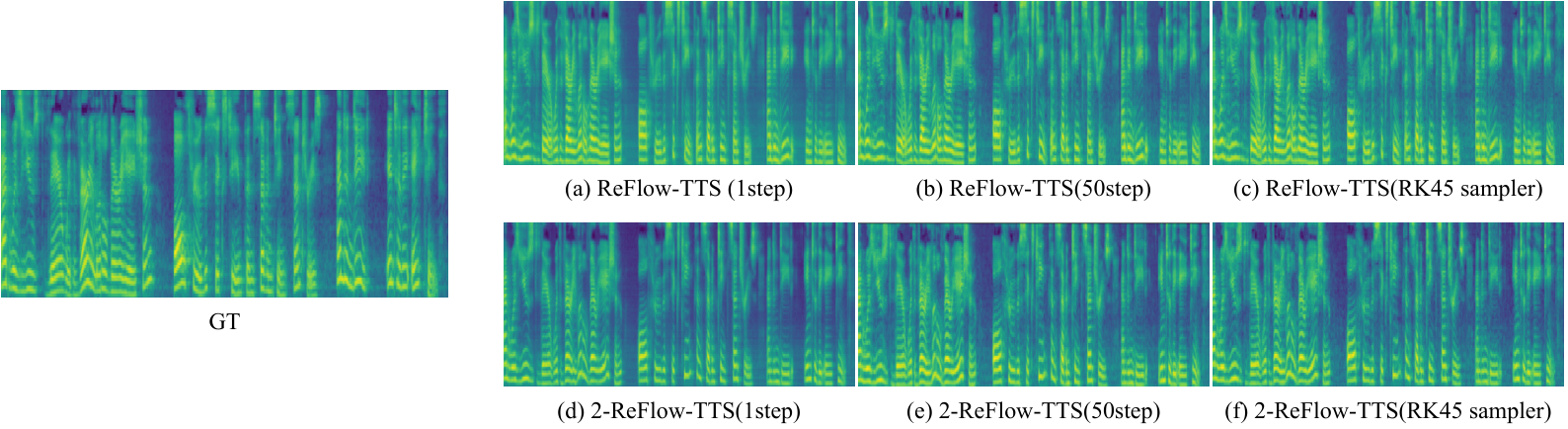}
  \caption{The visualization results of Mel-spectrograms for ReFlow-TTS and 2-ReFlow-TTS.}
  \label{fig:reflow_12th}
\end{figure*}

\begin{table}[!t]
	  \centering  
	  \caption{Evaluation results for ReFlow-TTS and 2-ReFlow-TTS.}
	  \label{t3} 
	  \begin{center}
	   \resizebox{8.5cm}{!}{
		\begin{tabular}{lcccc}\hline 
 		Method  & NFE $\downarrow$ & RTF $\downarrow$ & FD $\downarrow$ & MOS $\uparrow$  \\ \hline
 		ReFlow-TTS (1step)  & 1 & 0.00577 & 1.9872 & 4.16$\pm$0.09 \\
 		ReFlow-TTS (50step)  & 50 & 0.0964 & 0.2759 & 4.43$\pm$0.07 \\
 		ReFlow-TTS (RK45 solver) & 152 & 0.3703 & \textbf{0.1393} & \textbf{4.52$\pm$0.10} \\
 		\hline
 		2-ReFlow-TTS (1step) & 1 & \textbf{0.00562} & 2.0439 & 4.17$\pm$0.10 \\
 		2-ReFlow-TTS (50step) & 50 & 0.0886 & 0.3248 & 4.41$\pm$0.11 \\
 		2-ReFlow-TTS (RK45 solver) & 141 & 0.3669 & 0.2054 & 4.50$\pm$0.09 \\
 		\hline
		\end{tabular}}
	\end{center}
\end{table}

\vspace{-5pt}
\subsubsection{Comparative Models}
\vspace{-5pt}
We compare the four metrics mentioned above for the samples generated by the ReFlow-TTS and the following systems: 1) GT: This is the ground-truth recording; 2) GT (Mel + Voc): This is the speech synthesized using pretrained HiFi-GAN vocoder for GT Mel-spectrogram; 3) FastSpeech2~\cite{ren2020fastspeech}; 4) DiffGAN-TTS~\cite{liu2022diffgan}\footnote{\url{https://github.com/keonlee9420/DiffGAN-TTS}}; 5) ProDiff~\cite{huang2022prodiff}\footnote{\url{https://github.com/Rongjiehuang/ProDiff}}; 6) DiffSpeech~\cite{liu2022diffsinger}\footnote{\url{https://github.com/MoonInTheRiver/DiffSinger/blob/master/docs/README-TTS.md}}; 7) Grad-TTS~\cite{popov2021grad}\footnote{\url{ https://github.com/huawei-noah/Speech-Backbones/tree/main/Grad-TTS}}; 8) Diff-TTS~\cite{jeong21_interspeech}; 9) CoMoSpeech~\cite{ye2023comospeech}\footnote{\url{https://github.com/zhenye234/CoMoSpeech}}: we compare both the teacher model and the one-step CoMoSpeech with our proposed ReFlow-TTS model.
Note that the FastSpeech2 and Diff-TTS models are reproduced by ourselves, and the time step $T$ of Diff-TTS is set to 1000 for better performance.
\subsection{Audio Performance}

The evaluation results of TTS are shown in Table \ref{t1}. For audio quality, our ReFlow-TTS using RK45 ODE solver for inference achieves the highest MOS and the best FD scores among all methods. This demonstrates the superior performance on modeling data distribution of our proposed ReFlow-TTS.  Additionally, our ReFlow-TTS (1step) using only one sampling step achieves better performance than most previous diffusion based TTS models.
On the other hand, our ReFlow-TTS outperforms other models at NFE=50 while maintaining a significantly lower RTF. Compared to Diff-TTS, our model demonstrates superior performance even at a very low RTF. The detailed visualization results are demonstrated in Figure \ref{fig:comp_models}. Compared with Figure \ref{fig:comp_models} (h) and others, the Mel-spectrogram generated by ReFlow-TTS has richer details so that a more natural and expressive voice is produced.

Table \ref{t2} shows the evaluation results of TTS models using one sampling step. For audio quality, our ReFlow-TTS using one sampling step achieves competitive results without relying on a pre-trained Teacher model compared with the existing SOTA one-step diffusion model,  CoMoSpeech. Additionally, our ReFlow-TTS (1step) achieves lower RTF compared to CoMoSpeech. The detailed visualization results are demonstrated in Figure \ref{fig:exp_1step}.

We also conduct experiments for 2-ReFlow-TTS. The results are shown in Table \ref{t3}. It can be seen that under the premise of achieving similar results compared to ReFlow-TTS, the 2-ReFlow-TTS can get lower inference speed using Euler ODE solver for 1 step or 50 steps and RK45 ODE solver. This also demonstrates that the recursive rectified flow is more straight and easier to simulate numerically. The visualization results are demonstrated in Figure \ref{fig:reflow_12th}. Through the exploration of Reflow-TTS and 2-Reflow-TTS processes, we demonstrate the robustness of our proposed conditional rectified flow model.  Training only the Reflow-TTS model yields high-fidelity samples, eliminating the necessity for training a second rectified flow model.

\section{Conclusions}
In this paper, we propose a simple yet efficient ReFlow-TTS,  which is the first rectified flow model for speech synthesis. The ReFlow-TTS can synthesize speech samples with the best audio quality using an RK45 ODE solver for sampling. Furthermore, our proposed ReFlow-TTS by one step sampling can achieve better performance than most previous diffusion based TTS models and it does not rely on the pre-training of Teacher model for better performance. The ReFlow-TTS model significantly improves the audio quality and the  usability in real-world scenarios. The audio samples are publicly available at 
\url{https://gwh22.github.io/ReFlow-TTS/}.

\vfill\pagebreak
 \bibliographystyle{IEEEbib}
\bibliography{strings,refs}

\end{document}